\documentclass[pre,preprint,usletter,showpacs,superscriptaddress]{revtex4}

\usepackage{amsmath}
\usepackage{amsfonts}
\usepackage{amssymb}
\usepackage{color}
\usepackage{amsthm}
\usepackage [ansinew] {inputenc}
\usepackage{graphicx}
\usepackage{natbib}

\begin{document}
\title{Weakly coupled map lattice models for multicellular patterning and collective normalization of abnormal single-cell states}

\author{Vladimir Garc\'{\i}a-Morales}\email{garmovla@uv.es}
\author{Jos\'e A. Manzanares}
\author{Salvador Mafe}
\affiliation{Dept. de Termodin\`amica, Facultat de F\'{\i}sica, Universitat de Val\`encia, \\ E-46100 Burjassot, Spain}

\begin{abstract}
\noindent We present a weakly coupled map lattice model for patterning that explores the effects exerted by weakening the local dynamic rules on model biological and artificial networks composed of two-state building blocks (\emph{cells}). To this end, we use two cellular automata models based on: (i) a smooth majority rule (\emph{model I}) and (ii) a set of rules similar to those of Conway's \emph{Game of Life} (\emph{model II}). The normal and abnormal cell states evolve according with local rules that are modulated by a parameter $\kappa$. This parameter quantifies the effective weakening of the prescribed rules due to the limited coupling of each cell to its neighborhood and can be experimentally controlled by appropriate external agents. The emergent spatio-temporal maps of single-cell states should be of significance for positional information processes as well as for intercellular communication in tumorigenesis where the collective normalization of abnormal single-cell states by a predominantly normal neighborhood may be crucial.
 \end{abstract}
\maketitle

\section{Introduction}

Biological and artificial networks composed of micro- and nanoscale two-state building blocks (\emph{cells}) are bound to operate under significant physical limitations because of individual diversity and thermal noise effects. These effects may weaken the local dynamic rules of the network and result in a distribution of individual cell states instead of the two generic Boolean states 0 and 1. We present 2D, coupled map lattice models based on cellular automata dynamics \cite{JPHYSA} to explore the consequences of this weakening, with emphasis on biophysical multicellular ensembles. In \emph{model I}, the central cell state is determined by applying a smooth majority rule to the individual states of the multicellular neighborhood. In \emph{model II}, the dynamical rules are similar to those of Conway's \emph{Game of Life}  \cite{Berlekamp,Adamatzky,Adachi,Chua1,Pazienza}. The above rules may favor the normal state (0) but permit also the existence of cells in the abnormal state (1). The predominance of state 0 over state 1 may occur in the ensemble under certain conditions and is named \emph{normalization}.
 
For the two biologically-motivated models, the individual cell states evolve according with local dynamical rules modulated by a coupling parameter $\kappa$: This parameter quantifies the weakening of the rules due to the limited coupling of each individual cell to its local neighborhood. In general, low values of $\kappa$ tend to enforce the local rules over the ensemble while high values of $\kappa$ are associated with limited intercellular communication.
We note that $\kappa$ should have a wide physical significance. In the case of artificial networks, $\kappa$ could be related to the degree of heterogeneity characteristic of most nanostructures. For instance, nanowire field-effect transistors and nanoparticle-based single electron transistors do not show identical responses at the individual level because of significant physical variability and this experimental fact may result in weak collective coupling \cite{Cervera1}. Also, $\kappa$ can indirectly account for the decreased cooperativity observed in molecular monolayers because of thermal noise effects \cite{Cervera2}. In these cases, \emph{static} (individual variability) and \emph{dynamic} (finite temperature) noise limitations eventually result in \emph{weak local rules} of the system dynamics. Limited coupling may also be significant in clusters of protein ion channels with individually different threshold responses \cite{Cervera3} and interacting cells forming spatio-temporal patterns \cite{Cervera4}. These multicellular patterns are crucial to positional information processes such as embryogenesis and tumorigenesis \cite{Potter, Adams}. 
 
Abnormal tumour cells form pre-cancerous lesions that can rest dormant for a long time because they are located in unsuitable sites or controlled by intercellular interactions with a majority of normal cells \cite{Tarin, Sonnenschein}. In this context, exploring the consequences of the interaction between the abnormal tumor cells (state 1) and their neighboring normal cells (state 0) should be of interest to new theoretical approaches \cite{Baker, Sonnenschein2, Baker2, Rosenfeld}. In the \emph{tissue organization field theory} for instance, the disruption of local intercellular communication is associated with the appearance of abnormal cells and carcinogenesis \cite{Sonnenschein2, Baker2, Rosenfeld}. These facts suggest the possibility of acting on the intercellular coupling by means of appropriate external agents. However, strategies designed to modify multicellular ensembles are difficult to implement because small changes at the local level may result in unexpected global outcomes. Indeed, the emergent large-scale patterns cannot be easily anticipated from single-cell considerations \cite{Cervera4, Adams}. This problem is crucial in carcinogenesis where small clusters of cells may establish local interactions that escape from the morphogenetic control based on intercellular coupling \cite{Sonnenschein2,Baker,Rosenfeld}.

Modeling the interplay between the local rules that govern intercellular coupling and the emergent multicellular patterning is of current interest. We consider here two weakly coupled map lattice models for the spatio-temporal patterning and normalization of cell ensembles. In particular, we describe the range of single-cell states between 0 and 1 that may originate from the weakening of the intercellular local rules and show the dynamical consequences of weak coupling on multicellular patterning. 

\section{Models}

The states of biological cells can be defined in terms of dynamical variables such as the concentration $c$ of a signaling molecule in the cell \cite{Potter, Trosko} and the membrane potential $V$, defined as the electric potential difference between the cell cytoplasm and the extracellular microenvironment under zero current conditions \cite{Cervera4, Chernet}. In general, a variable $x$ that characterizes the cell state can be mapped into a dimensionless variable $u$ varying in the range $[0, 1]$ by the transformation $u = (x -x_0)/(x_1 -x_0)$, where $x_0$ and $x_1$ are, respectively, the values of $x$ in some reference normal (0) and abnormal (1) states. The membrane potential $V$ constitutes a typical example of dynamical variable because depolarized potentials are characteristic of abnormal cells \cite{Chernet,Yang,Levin1,Levin2}. The corresponding dimensionless variable would be $u = (V - V_0)/(V_1 - V_0)$ where $V<0$, with $V_0$ and $V_1$ the normal (polarized) and abnormal (depolarized) potentials $(V_0 \le V \le V_1)$. The variable $u$ varies continuously between 0 and 1 and characterizes the cell state. We model the multicellular ensemble as a 2D lattice where each site represents a single cell. The site can be in a continuum of states ranging from 0 (normal state) to 1 (abnormal state). A lattice where most sites are found in state 0 is said to be in a normalized state. The biological signals that couple individual cells to their local multicellular environment may contribute to  normalization and are modeled using a continuous parameter $\kappa$ that accounts for a weak coupling between cells.

~\\
\textbf{Lattice and states} ~\\

We consider a 2D square lattice $L$ with square (Moore) neighborhoods of $3\times 3$ sites. The neighborhood of the lattice site $(i,j)$ is formed by the sites $(i+k,j+m)$ where $k$ and $m$ can take the values $-1,\ 0$ and $1$. At time $t$, the state of the site $(i,j)$ is given by the continuous dimensionless dynamical variable $u_t^{i,j} \in [0,1]$. The states of all sites in the lattice are synchronously updated at discrete time steps according to the map
\begin{equation}
u_{t+1}^{i,j}=f(u_{t}^{i,j}, s_{t}^{i,j}; \kappa) \qquad \qquad (i,j) \in L \label{mapper}
\end{equation}
where $\kappa$ is the coupling parameter and 
\begin{equation}
s_{t}^{i,j}\equiv \sum_{k,m=-1}^{1}u_{t}^{i+k,j+m} 
\end{equation}
is the neighborhood sum.

The lattice dynamics is studied using numerical simulations and the analytical mean-field approximation
\begin{equation}
u_{t+1} = f(u_{t}, 9u_{t}; \kappa)\equiv f_{\text{\text{MF}}}(u_{t}; \kappa) \label{mapperMF}
\end{equation}
which considers that all cells have approximately the same average value
\begin{equation}
u_{t}=\left<u_{t}^{i,j}\right>\equiv \frac{1}{\Omega}\sum_{i=1}^{n}\sum_{j=1}^{n}u_{t}^{i,j}
\end{equation}
where $\Omega=n^{2}$ is the total number of sites, with $n$ the number of sites on a side of the square lattice. The mean field approximation provides a good description of the dynamics, Eq. (\ref{mapper}), only if
\begin{eqnarray}
\left<f(u_{t}^{i,j}, s_{t}^{i,j}; \kappa)\right>&\approx& f(\left<u_{t}^{i,j}\right>, \left<s_{t}^{i,j}\right>; \kappa)= f_{\text{\text{MF}}}(u_{t}; \kappa) \label{MFmp}
\end{eqnarray}
Therefore, by expanding Eq. (\ref{mapper}) around $u_{t}^{i,j}=u_{t}$, we obtain
\begin{equation}
u_{t+1}^{i,j}= f_{\text{\text{MF}}}(u_{t}; \kappa)+(u_{t}^{i,j}-u_{t})\left.\frac{\partial f}{\partial u_{t}^{i,j}}\right|_{u_{t}^{i,j}=u_{t}}+(s_{t}^{i,j}-9u_{t})\left.\frac{\partial f}{\partial s_{t}^{i,j}}\right|_{u_{t}^{i,j}=u_{t}}
+\ldots \label{superexpan}
\end{equation}
Thus, the validity of the mean-field approximation as a reduced description of the full dynamics depends on (a) the convergence of this series and (b) the fulfillment of the relationship
\begin{equation}
\left|\left<(u_{t}^{i,j}-u_{t})\left.\frac{\partial f}{\partial u_{t}^{i,j}}\right|_{u_{t}^{i,j}=u_{t}}+(s_{t}^{i,j}-9u_{t})\left.\frac{\partial f}{\partial s_{t}^{i,j}}\right|_{u_{t}^{i,j}=u_{t}}\right>   \right| <<  f_{\text{\text{MF}}}(u_{t}; \kappa) \label{supercrite}
\end{equation}

Specific models of Eq. (\ref{mapper}) are constructed following the method of Ref. \cite{JPHYSA}. First, the limits $\kappa \to 0$ and $\kappa \to \infty$ of Eq. (\ref{mapper}), named the cellular automata limits, are described in terms of simple rules. Then, we allow $\kappa$ to take any finite non-zero value, thus weakening the dynamics of the cellular automata limits.

~\\
\textbf{Model I}~\\

This model constitutes a \emph{smooth majority} coupled map lattice. The coupling between sites is modulated by the continuous parameter $\kappa \in (0,\infty)$:   
\begin{itemize}
\item   \emph{In the limit $\kappa\to 0$ the central site within a neighborhood remains in state $1$ at the next time step if and only if there are no less than $8$ other neighboring sites in state $1$ as well. Otherwise, it changes to state $0$ at the next time step.} In this limit, the only possibility for a cell to remain abnormal is that all cells in the multicellular ensemble are abnormal; otherwise, normalization of the ensemble occurs after a transient. 
\item   \emph{In the limit $\kappa \to \infty$ a site remains in state $0$ at the next time step if and only if all neighboring sites are in state $0$ as well. Otherwise, it changes to state $1$ at the next time step.} This limiting case corresponds to the 
predominance of the abnormal state in absence of intercellular coupling, the situation opposite to the above case $\kappa \to 0$.
\end{itemize}
At intermediate values of $\kappa$, sites with states 0 and 1 should coexist in the ensemble. The above cases are the \emph{cellular automaton limits} of the coupled map lattice. Note that in these limits a site with state 0 (respectively 1) cannot arise at the center of a neighborhood where all sites have value 1 (respectively 0). Therefore, homogeneous neighborhoods where all sites have value either 0 or 1 are fixed points of the dynamics.

According with these rules, the map of Eq. (\ref{mapper}) that governs the spatio-temporal evolution of the state $u_{t}^{i,j}$ of cell $(i,j)$ is given by: 
\begin{equation}
u_{t+1}^{i,j}=\frac{\mathcal{B}_{\kappa}\left(9-s_{t}^{i,j},\ \frac{1}{2}\right)\mathcal{B}_{1/\kappa}\left(5-s_{t}^{i,j},\ \frac{9}{2}\right)}{\mathcal{B}_{\kappa}\left(0,\ \frac{1}{2}\right)\mathcal{B}_{1/\kappa}\left(0,\ \frac{9}{2}\right)} \label{majok}
\end{equation}
where the $\mathcal{B}_{\kappa}$-function of real variables $x$ and $y$ is \cite{JPHYSA}:
\begin{equation}
\mathcal{B}_{\kappa}(x,y)\equiv \frac{1}{2}\left[ 
\tanh\left(\frac{x+y}{\kappa} \right)-\tanh\left(\frac{x-y}{\kappa} \right) \right]
 \label{bkappa}
\end{equation} 
Note that $\kappa$ is the only free parameter of the model and modulates the local rules that couple the multicellular ensemble. For all finite values of the real variables $x$ and $y$, the $\mathcal{B}_{\kappa}$-function satisfies the limits \cite{JPHYSA}:
\begin{eqnarray}
\lim_{\kappa \to \infty}\mathcal{B}_{\kappa}\left(x,\ y\right)&=& 0 \qquad  \qquad
\lim_{\kappa \to \infty}\frac{\mathcal{B}_{\kappa}\left(x,\ y\right)}{\mathcal{B}_{\kappa}\left(0,\ y\right)}=1 \label{lim1} \\
\lim_{\kappa \to 0}\mathcal{B}_{\kappa}\left(x,\ y\right)&=& \mathcal{B}(x,y)=\frac{1}{2}\left(\frac{x+y}{|x+y|}-\frac{x-y}{|x-y|}\right)={\begin{cases} \text{sgn}\ y &{\text{if }} |x| < |y|\\ \frac{\text{sgn}\ y}{2}&{\text{if }}|x|=|y| \\0&{\text{if }}|x| > |y| \end{cases}} \label{lim2}
\end{eqnarray}
where we have introduced the $\mathcal{B}$-function, $\mathcal{B}(x,y)$, which allows a universal map for cellular automata to be formulated \cite{VGM1}. 

In the limit $\kappa \to 0$, Eq. (\ref{majok}) becomes 
\begin{equation}
u_{t+1}^{i,j}=\mathcal{B}\left(9-s_{t}^{i,j},\ \frac{1}{2}\right) \label{limeq0}
\end{equation}
For initial conditions that satisfy
\begin{equation}
s_0^{i,j}\in \mathbb{R} \setminus (\mathbb{Z}/2) \quad \forall  \ (i,j) \in L \label{condition}
\end{equation}
where $\mathbb{R} \setminus (\mathbb{Z}/2)$ denotes the real line excluding all half integers $\frac{n}{2}$, $n\in \mathbb{Z}$, the first iteration of Eq. (\ref{limeq0}) becomes locally a map $\mathbb{R} \setminus (\mathbb{Z}/2) \to \mathcal{A}_{2}\equiv \{0,1\}$. From the second iteration, it collapses to a map $\left[\mathcal{A}_{2}\right]^9 \to \mathcal{A}_{2}$ where $[\mathcal{A}_{2}]^9$ is the Cartesian product of $9$ copies of the Boolean set $\mathcal{A}_{2}\equiv \{0,1\}$. Thus, for $t\ge 1$, Eq. (\ref{limeq0}) corresponds to the 2D totalistic Boolean cellular automaton \cite{VGM1, Ilachinski} that sets $u_{t+1}^{i,j}=1$ if $s_{t}^{i,j}=9$ and $u_{t+1}^{i,j}=0$ otherwise. It is to be noted that this cellular automaton behavior is found for most initial conditions since those that fail to satisfy Eq. (\ref{condition}) constitute a set with zero measure. Therefore, in the limit $\kappa \to 0$ and $t\ge 1$ a configuration with non-integer neighborhood sum $s_t^{i,j}$ cannot arise in the spatiotemporal dynamics if all neighborhoods satisfy Eq. (\ref{condition}) at $t=0$.

In the limit $\kappa \to \infty$, Eq. (\ref{majok}) becomes
\begin{equation}
u_{t+1}^{i,j}=\mathcal{B}\left(5-s_{t}^{i,j},\ \frac{9}{2}\right)\label{limeq1}
\end{equation}
which, for initial conditions that satisfy Eq. (\ref{condition}), collapses for $t\ge 1$ to a 2D totalistic cellular automaton that sets $u_{t+1}^{i,j}=0$ if $s_{t}^{i,j}=0$ and $u_{t+1}^{i,j}=1$ otherwise. Symmetry considerations \cite{VGM2, VGM3} show that Eq. (\ref{limeq1}) is the global complement of Eq. (\ref{limeq0}) so that the respective evolutions of these equations are the `negative' of each other if one exchanges normal and abnormal cells (see \emph{Appendix}). 

\begin{figure*} 
\begin{center}
\includegraphics[width=1.0 \textwidth]{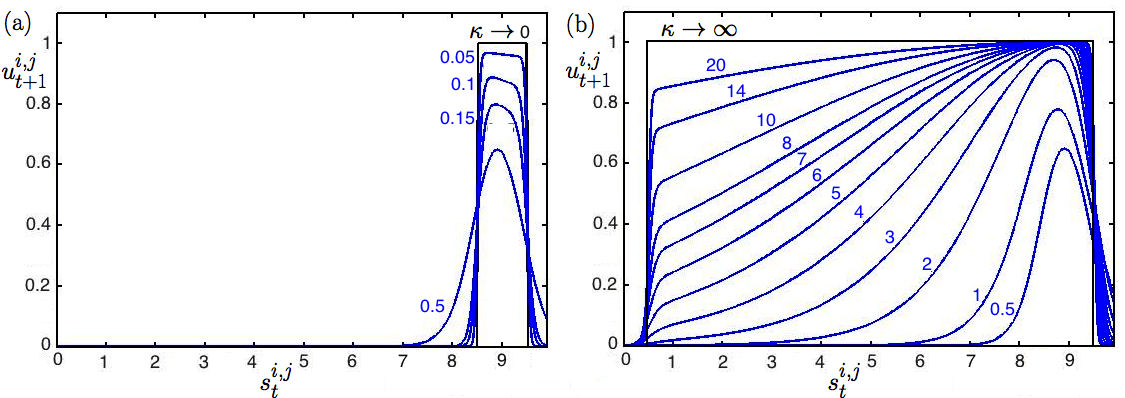}
\caption{\scriptsize{The cell state $u_{t+1}^{i,j}\in [0,1]$ vs. the neighborhood sum $s_{t}^{i,j}$ for the values of $\kappa$ indicated besides the curves. The panels are separated to better show the effect of varying $\kappa$ in Eq. (\ref{majok}). Note that values of $s_{t}^{i,j}>9$ are considered only to  show the mathematical trends of Eq. (\ref{majok})}} \label{model2f}
\end{center} 
\end{figure*}

Fig. \ref{model2f} shows $u_{t+1}^{i,j}$ of Eq. (\ref{majok}) as a function of the neighborhood sum $s_{t}^{i,j}$ for different values of $\kappa$. Note that normalization is enforced in the limit of low $\kappa$, Fig. \ref{model2f} (a), because a site in state 1 is viable only if all neighboring sites are in a state close to 1. On the contrary, normalization is discouraged in the limit of high $\kappa$, Fig. \ref{model2f} (b), because a site in state 0 requires that all neighboring sites are previously in a state close to 0. 
~\\

\textbf{Model II}~\\

In the limit $\kappa \to 0$, the rules of this model are similar to Conway's \emph{Game of Life} \cite{Berlekamp, Adamatzky, Adachi} for initial conditions $u_0^{i,j}=0$ or $1$, $\forall (i,j) \in L$: 
\begin{itemize}
\item   1. \emph{Any site in state $1$ with fewer than two nearest neighbors in state $1$ takes state $0$ at the next time step.} The rule establishes the normalizing effect of the local neighborhood when normal cells predominate.
\item   2. \emph{Any site in state $1$ with two or three nearest neighbors in state $1$ remains in state $1$ at the next time step.} The rule assumes that the normalization effect of the local neighborhood is lost when sufficient abnormal cells are present. 
\item   3. \emph{Any site in state $0$ with three nearest neighbors in state $1$ changes to state $1$ at the next time step.} The rule considers the promotion from a normal to an abnormal state.
\item   4. \emph{Any site in state $1$ with more than three nearest neighbors in state $1$ changes to state $0$ at the next time step.} The rule establishes a limit to abnormal cell expansion, e.g. because of finite available resources, representing a change from positive to negative cooperativity.  
\end{itemize}
For $u_{t}^{i,j}$ constrained to 0 or 1, these rules constitute the popular \emph{Game of Life} \cite{VGM1}, an outer totalistic cellular automaton discovered by Conway \cite{Berlekamp}, 
\begin{equation}
u_{t+1}^{i,j}={\begin{cases} u_{t}^{i,j} &{\text{if }} s_{t}^{i,j}-u_{t}^{i,j}=2 \\ 1 &{\text{if }} s_{t}^{i,j}-u_{t}^{i,j}=3    \\ 0&{\text{otherwise }} \end{cases}} \label{Lifeklim0}
\end{equation}
It can be proved (see the Appendix) that Eq. (\ref{Lifeklim0}) is equivalent to 
\begin{equation}
u_{t+1}^{i,j}={\begin{cases} \text 1 &{\text{if }} s_{t}^{i,j}=3 \\ u_{t}^{i,j} &{\text{if }} s_{t}^{i,j}=4    \\ 0&{\text{otherwise }} \end{cases}} \label{Lifeklim}
\end{equation}
However, we consider a more general 'fuzzy' dynamics controlled by a modulating parameter $\kappa$ which is finite and non-vanishing, with $u_t^{i,j} \in [0,1]$ a continuous variable. Further, we have:
\begin{itemize}
\item 5. \emph{The coupling between sites due to the above local rules is modulated by the parameter $\kappa \in (0,\infty)$}. This parameter loosely incorporates the collective influence of biological phenomena such as the stochastic intercellular diffusion of signaling molecules, the intrinsically probabilistic gene expression, and the individual cell heterogeneity. These noisy phenomena should weaken rules 1 to 4 above, which hold exactly only in the limit $\kappa \to 0$.  
\end{itemize}

Note that a predominantly normal neighborhood may constitute a normalizing microenvironment for a cell because of the abnormal cell underpopulation (rule 1). On the contrary, a significantly abnormal neighborhood may impair the normalization effect and promote the abnormal state (rules 2 and 3). In the case of abnormal cell overcrowding, however, limited proliferation could arise because of the competition for finite resources (rule 4).

All rules above are concisely implemented using the following map for the spatio-temporal evolution of $u_{t}^{i,j}$ in Eq. (\ref{mapper}):
\begin{equation}
u_{t+1}^{i,j}=\mathcal{B}_{\kappa}\left(3-s_{t}^{i,j},\ \frac{1}{2}\right)+u_{t}^{i,j}\mathcal{B}_{\kappa}\left(4-s_{t}^{i,j},\ \frac{1}{2}\right) \label{Lifek}
\end{equation}
In the limit $\kappa \to 0$, for initial conditions that satisfy Eq. (\ref{condition}) and such that no $u_{0}^{i,j}$ is in the interval $[0.4,0.5]$, Eq. (\ref{Lifek}) coincides with the \emph{Game of Life} cellular automaton of Eq. (\ref{Lifeklim}) since the variable $u_{t}^{i,j}$ becomes Boolean for $t\ge 1$. For $\kappa \ne 0$, the dynamics becomes fuzzy and the values of $u_{t}^{i,j}$ for $t\ge 1$ are bounded above by
\begin{equation}
u_{max}=\mathcal{B}_{\kappa}\left(0,\ \frac{1}{2}\right)+\mathcal{B}_{\kappa}\left(1,\ \frac{1}{2}\right)=
 \mathcal{B}_{\kappa}\left(\frac{1}{2},\ 1\right) \lesssim 1 \label{boundbelow}
\end{equation}
as it is obtained by replacing $u_{t}^{i,j}=1$ and $s_{t}^{i,j}=3$ or $4$ in Eq. (\ref{Lifek}) and using the block coalescence property of the $\mathcal{B}_{\kappa}$-function \cite{JPHYSA}. The lower bound
\begin{equation}
u_{min}=\mathcal{B}_{\kappa}\left(5,\ \frac{1}{2}\right)\gtrsim 0 \label{boundabove}
\end{equation}
is obtained by replacing $u_{t}^{i,j}=0$ and $s_{t}^{i,j}=8$ in Eq. (\ref{Lifek}). Thus, $u_{t}^{i,j}$ is constrained to a subset of the unit interval $u_{t}^{i,j}\in [u_{min},u_{max}]$ determined by $\kappa$.

\section{Results and Discussion}

We have carried out numerical simulations with \emph{models I} and $II$ assuming periodic boundary conditions. We consider first a generic initial condition consisting of a random distribution of 0 and 1 states with density approximately equal to $0.5$.

\subsection{Model I}

\begin{figure} 
\begin{center}
\includegraphics[width=1.0 \textwidth]{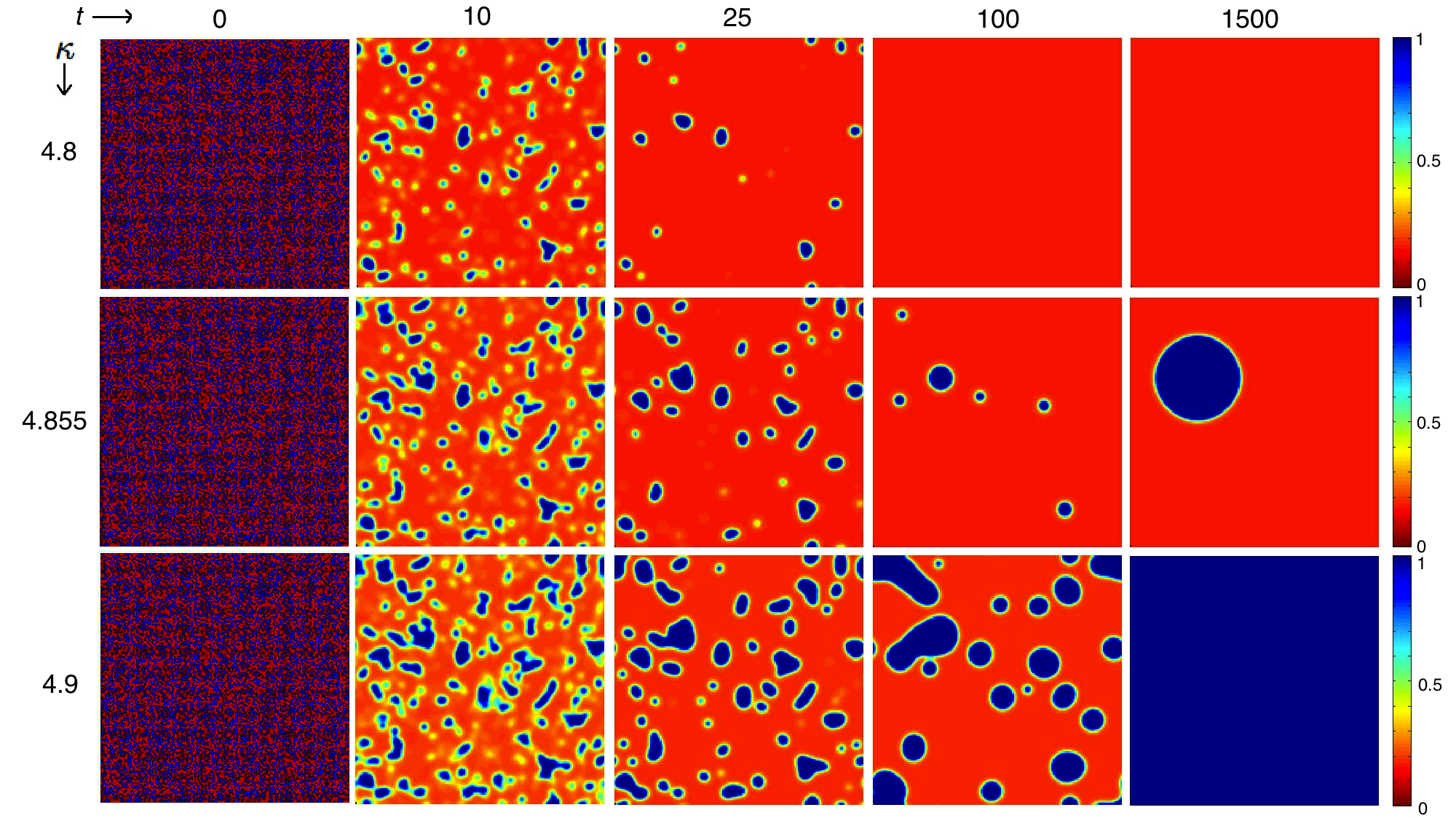}
\caption{\scriptsize{(Color online) Spatio-temporal evolution of the cell states $u_{t}^{i,j}$ taking values between 0 and 1 (right bar) for \emph{model I} obtained by iterating Eq. (\ref{majok}) in a multicellular ensemble of $159 \times 159 = 25281$ cells at different times $t$ for three $\kappa$ values. The initial ($t = 0$) state with cells randomly distributed in the 0 and 1 states is the same for the three cases.}} \label{patmajo}
\end{center} 
\end{figure}

\begin{figure*} 
\begin{center}
\includegraphics[width=0.6 \textwidth]{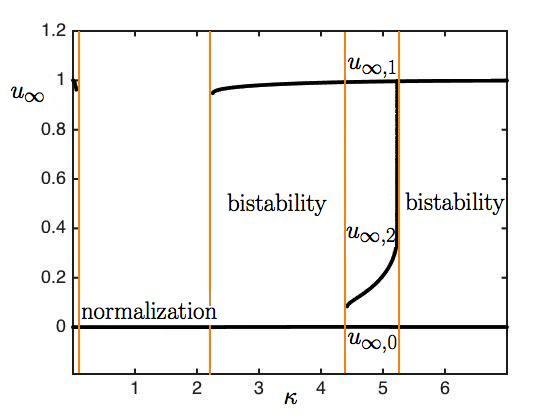}
\caption{\scriptsize{(Color online) Bifurcation diagram calculated from the asymptotic behavior of Eq. (\ref{majobif}). The stationary cell state $u_{\infty}$ obtained after $3\cdot 10^4$ time steps is shown as a function of the parameter $\kappa$. }} \label{BIFno}
\end{center} 
\end{figure*}

Fig. \ref{patmajo} shows snapshots at different dimensionless times of the multicellular ensemble evolution determined by Eq. (\ref{majok}). After a sufficiently long time, the system reaches a homogeneous state that can be either normal (upper panels) or abnormal (bottom panels). The duration of the transient leading to homogeneity depends on the distance to the transition separating these two trends (see the bifurcation diagram in Fig. \ref{BIFno}). The reversion of abnormal (blue color) to normal (red color) cell states is only possible for low enough values of $\kappa$, which promote a correction mechanism of the locally abnormal pattern at $t = 0$. Indeed, the weakening of the local rules favoring the normal state occurs at high enough values of $\kappa$. This fact causes the expansion of the abnormal state as $\kappa$ is increased above $\kappa \approx 4.9$. 

The above results can be understood if we reduce the map Eq. (\ref{majok}) to the case of homogeneous neighborhoods. This corresponds to the mean field approximation, Eq. (\ref{mapperMF}). Within this approximation, all neighborhoods in Eq. (\ref{majok}) are decoupled and the labels $i,j$ can be dropped because we are describing an average single-cell behavior, thus taking $u_{t}^{i+k,j+m}=u_{t}$ for all $k, m \in \{-1,0,1\}$ and $s_{t}^{i,j}=9u_{t}$. (Note that, because of dynamical fluctuations, the local value of the general dynamics may depart from this single-cell mean-field value.) This coarse-grained approximation is useful for capturing the dynamics because no inhomogeneous neighborhoods can persist in the cellular automata limits of the model. The mean field approximation of Eq. (\ref{majok}) is
\begin{equation}
u_{t+1}=\frac{\mathcal{B}_{\kappa}\left(9-9u_{t},\ \frac{1}{2}\right)\mathcal{B}_{1/\kappa}\left(5-9u_{t},\ \frac{9}{2}\right)}{\mathcal{B}_{\kappa}\left(0,\ \frac{1}{2}\right)\mathcal{B}_{1/\kappa}\left(0,\ \frac{9}{2}\right)} \label{majobif}
\end{equation}
The bifurcation diagram of this map (Fig. \ref{BIFno}) provides the stable fixed points that can be dynamically reached depending on the initial condition. For constructing the diagram, the whole interval of initial conditions $u_{0}\in [0,1]$ is sampled and the dynamics is then iterated to calculate $u_{\infty}$. Bistable regimes are found for several parameter ranges. In the range  $4.4 \lesssim \kappa \lesssim 5.2$, three stable states coexist and depending on the initial conditions, the system can converge either to the normal state $u_{\infty,0} \approx 0$, to the abnormal state $u_{\infty,1} \approx 1$ (most prominent at high $\kappa$ values) or to a third stable intermediate state $u_{\infty,2}$ found only at an intermediate $\kappa$ regime. For the particular initial condition of Fig. \ref{patmajo}, the critical value $\kappa \approx 4.855$ marks the transition between the attractor corresponding to $u_{\infty,2}$ (light regions in the rightmost panels of Fig. \ref{patmajo}) and that corresponding to $u_{\infty,1}$ (dark regions). For larger $\kappa$, the abnormal state is the most prominent, attracting almost all trajectories in phase space. The above mean-field analysis is independent of the total number of cells $\Omega$ in the ensemble. Numerical simulations of \emph{model I}, Eq. (\ref{majok}) showed that the mean field approximation accurately captures its average spatiotemporal dynamics: all fixed points correspond to homogeneous states and the series expansion, Eq. (\ref{superexpan}), converges as Eq. (\ref{supercrite}) is satisfied by most trajectories. Even at intermediate values of $\kappa$, where curved and circular interfaces are observed (see Fig. \ref{patmajo}), the numerical simulations showed that the temporary contribution of the cells at domain interfaces can be neglected compared to the  dominant bulk domains that contain most of the sites. This amounts to neglect the contribution to the lattice average of the small fraction of neighborhoods for which Eq. (\ref{supercrite}) does not hold because the derivatives in that equation are large, i.e. the contribution of those sites found at the curvy and circular interfaces separating the more prominent bulk domains (which have a dynamical state corresponding to the different fixed points of the mean field approximation).

\begin{figure*} 
\begin{center}
\includegraphics[width=0.6 \textwidth]{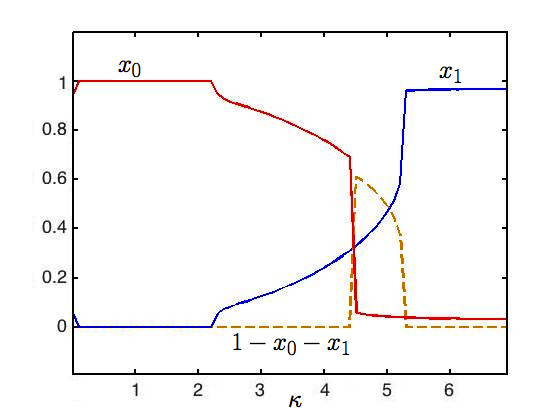}
\caption{\scriptsize{(Color online) Fraction of initial conditions in the unit interval that are attracted as $t\to \infty$ to the stable state 0 ($x_{0}$ curve) and to state 1 ($x_{1}$ curve) as a function of $\kappa$ and a third stable state $u_{\infty,2}$ found at intermediate $\kappa$ values ($1-x_0-x_1$ dashed curve).}} \label{sts}
\end{center} 
\end{figure*}

Further insight is obtained by estimating of the width of the basin of attraction for the different fixed points. The fractions $x_{0}$, $x_{1}$ and $x_{2}=1-x_{0}-x_{1}$ of the initial conditions in the unit interval attracted by the fixed points $u_{\infty,0}$, $u_{\infty,1}$ and $u_{\infty,2}$, respectively, are plotted in Fig. \ref{sts}. The homogeneous normal state $u_{\infty,0} \approx 0$ is most prominent at low $\kappa$ values, the abnormal state $u_{\infty,1} \approx 1$ dominates at high $\kappa$ values, and the fixed point $u_{\infty,2}$ is only found at intermediate $\kappa$, in significant competition with the abnormal state $u_{\infty,1} \approx 1$. It is in this intermediate range of $\kappa$ where the bubbles shown in Fig. \ref{patmajo} persist during long time spans. Therefore, normalization can be achieved by lowering $\kappa$ in ensembles where abnormal cells dominate.

Experimentally, the initial cancer stages have been associated with limited or defective intercellular communication in multicellular ensembles \cite{Tarin, Sonnenschein2, Rosenfeld, Banerjee, Mesnil, Schalper}. As expected, Fig. \ref{patmajo} suggests that restoring the intercellular coupling (i.e. lowering the value of $\kappa$) by means of external agents could contribute to ensemble normalization. However, the effects of this restoring procedure depend on the local rules and the particular initial conditions, as we show in the next model.

\subsection{Model II}

Imagine a multicellular ensemble with $\kappa$ finite and a small number of abnormal cells at $t=0$. Because the \emph{Game of Life} rules are exact in the limit $\kappa \to 0$, full normalization can no longer be warranted in \emph{model II}. Indeed, the \emph{Game of Life} displays complex behavior for generic initial conditions and, hence, abnormal cells could persist. Furthermore, lowering $\kappa$ from a sufficiently high value of this parameter may even enhance the contribution of the abnormal cells to the total ensemble for certain particular cases.

\begin{figure} 
\begin{center}
\includegraphics[width=0.98 \textwidth]{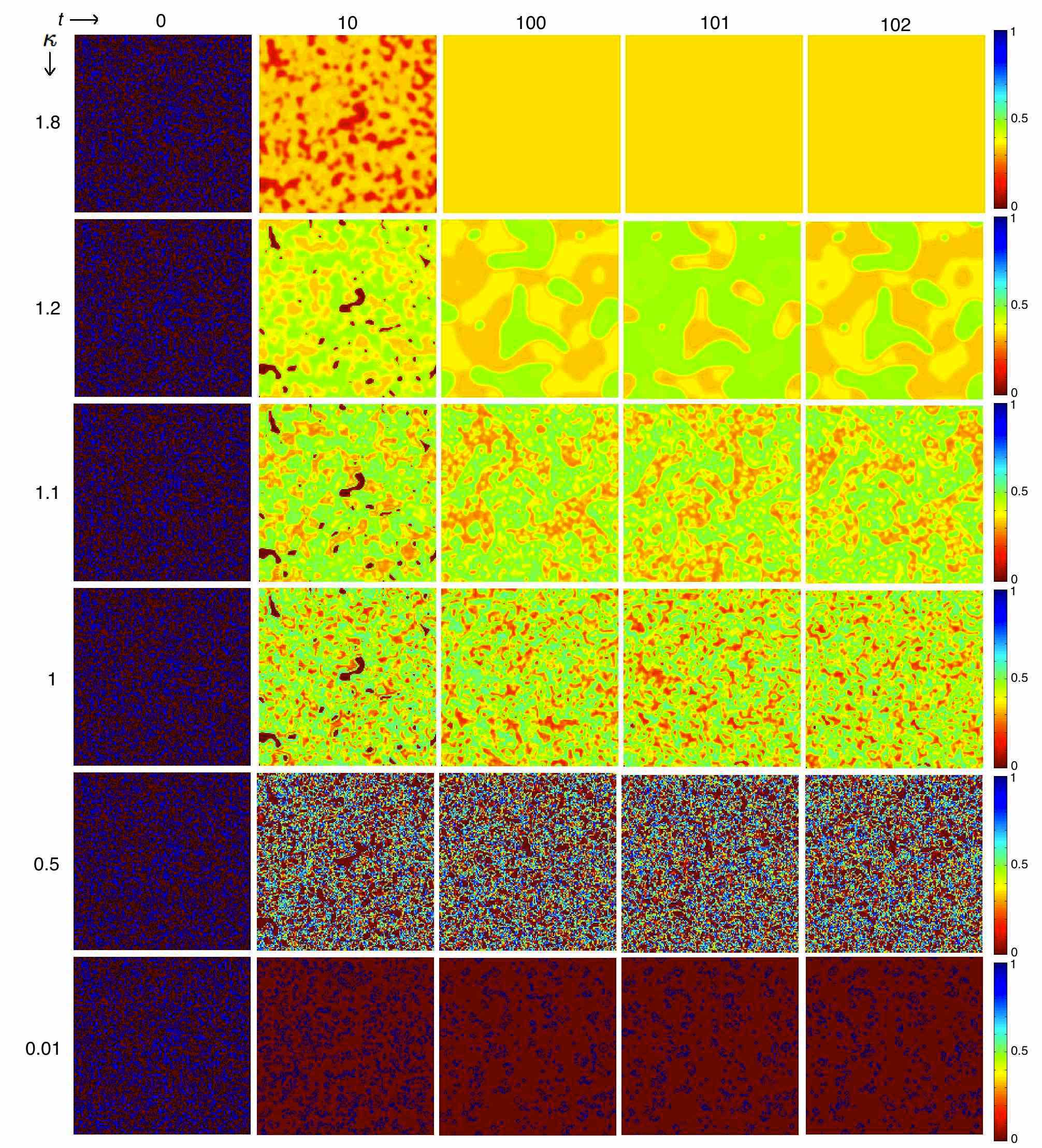}
\caption{\scriptsize{(Color online) Spatio-temporal evolution of the cell states $u_{t}^{i,j}$ taking values between 0 and 1 (right bar) for \emph{model II} obtained by iterating Eq. (\ref{Lifek}) in a multicellular ensemble of $159 \times 159 = 25281$ cells for six different $\kappa$ values. The initial ($t = 0$) state with cells randomly distributed in the 0 and 1 states is the same for all cases.}} \label{overview}
\end{center}
\end{figure}

Fig. \ref{overview} shows the snapshots of the multicellular ensemble for \emph{model II}, Eq. (\ref{Lifek}), at different times. For $\kappa$ sufficiently large, the system reaches, after a transient, a homogeneous state  that appears to be only slightly abnormal. However, as $\kappa$ is lowered, a bifurcation to oscillatory behavior is observed for domains of abnormal cells. Decreasing $\kappa$ further, the number of oscillatory components is increased and the system exhibits a transition to strongly aperiodic behavior, that is most prominent when $\kappa = 1$. 
For $\kappa <1$ the patterns are noisy and the cell state $u_t^{i,j}$ varies continuously with time within the interval $[u_{min}, u_{max}]$ given by Eqs. (\ref{boundbelow}) and (\ref{boundabove}). However, the intermediate states collapse as $\kappa \to 0$  and the cells show only the discrete states 0 and 1. In this limit, the dynamics reduces to the \emph{Game of Life}. For generic initial conditions, therefore, the ensemble may fail to normalize when $\kappa$ is decreased from a particular value.

\begin{figure} 
\begin{center}
\includegraphics[width=1.0 \textwidth]{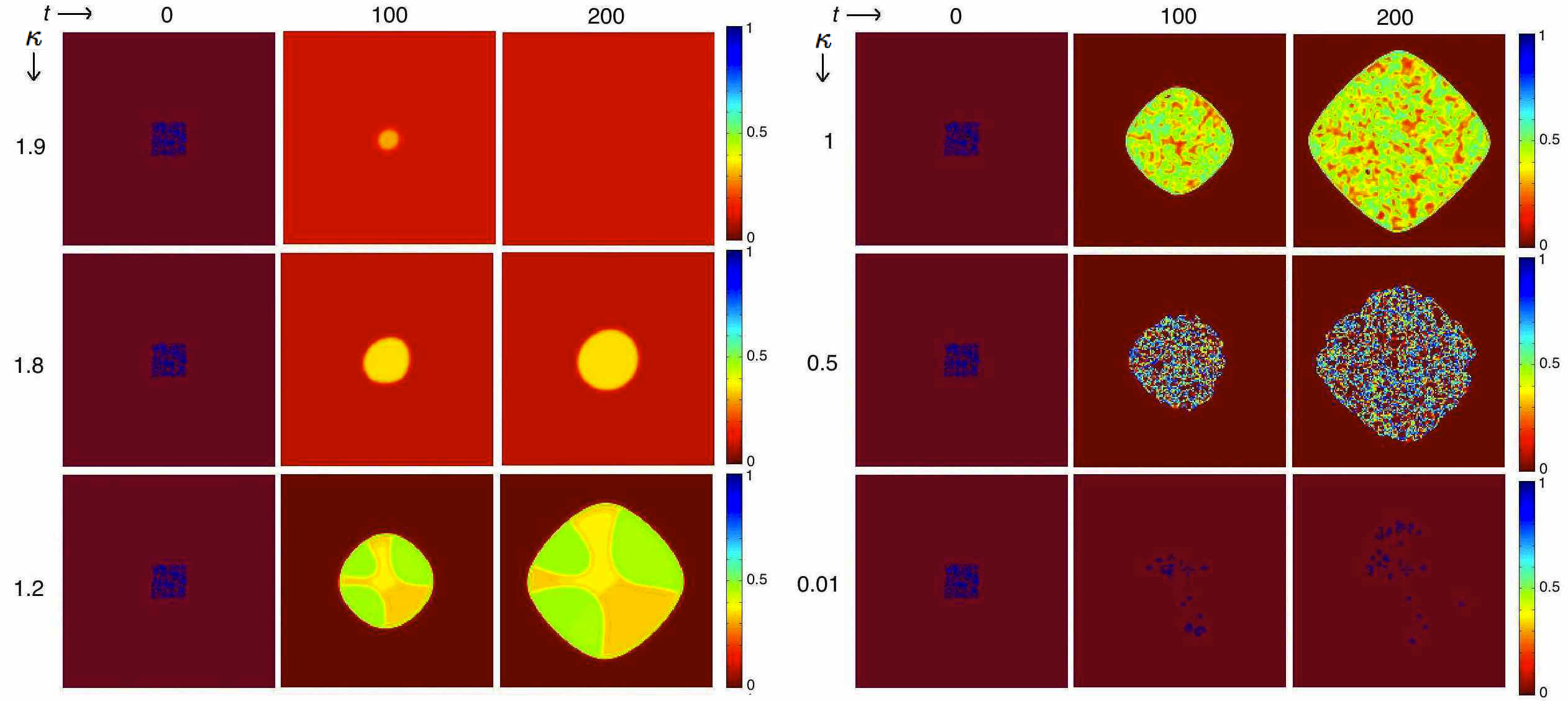}
\caption{\scriptsize{(Color online) Spatio-temporal evolution of the cell states $u_{t}^{i,j}$ taking values between 0 and 1 (right bar) for \emph{model II} obtained by iterating Eq. (\ref{Lifek}) in a multicellular ensemble of $159 \times 159 = 25281$ cells for six different $\kappa$ values. The initial ($t = 0$) state is the same for all cases and consists of a random distribution of 0 and 1 values in a central square region of the lattice of $25\times 25$ size, the rest of the lattice being at state 0.}} \label{overview2}
\end{center}
\end{figure}

To emphasize the complexity of the ensemble normalization,
Fig. \ref{overview2} shows the snapshots obtained for an inhomogeneous region occupying initially a central cluster. If $\kappa \lesssim 1.9$ a homogeneous normal state is obtained at long times. As $\kappa$ is decreased, the central inhomogeneity can grow.  Domain formation and oscillations are observed within the growing inhomogeneity (see also Fig. \ref{overview}). 

\begin{figure*} 
\begin{center}
\includegraphics[width=1.0 \textwidth]{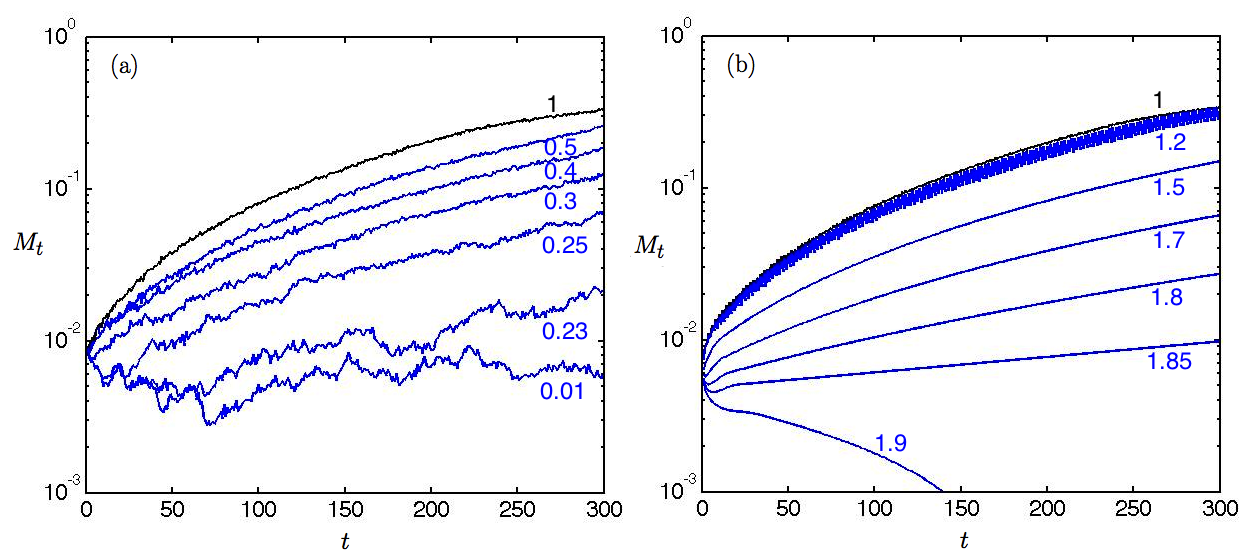}
\caption{\scriptsize{(Color online) $M_t$ versus time $t$ obtained from Eqs. (\ref{Lifek}) and (\ref{mass}) for the same initial condition as in Fig. \ref{overview2} and the values of $\kappa$ indicated on the curves. $M_{t}$ constitutes a measure of the ensemble abnormality.}} \label{mascu} 
\end{center}
\end{figure*}

To better understand the results of Fig. \ref{overview2}, let $w_{t}\approx 0$ denote the state $u_{t}^{i,j}$ of a cell in the homogeneous region of the ensemble far away from the inhomogeneity.  Then, the time-dependent variable
\begin{equation}
M_t=\frac{1}{\Omega}\sum_{i=1}^{n}\sum_{j=1}^{n}u_{t}^{i,j}-w_{t}=u_{t}-w_{t} \label{mass}
\end{equation}
provides an estimate of the relative weight of abnormal cells in the lattice with respect to $w_t$. Fig. \ref{mascu} shows $M_t$ calculated from Eqs. (\ref{Lifek}) and (\ref{mass}) and the same initial condition as in Fig. \ref{overview2}. For $\kappa=1$, the optimal growth of the abnormal region is obtained. The impact of the domain oscillations within the abnormal region is clearly visible for $\kappa=1.2$. The effects of noise are more prominent as $\kappa <1$ is decreased. Statistically, fluctuations are more noticeable when addition is performed over the values  $u_{t}^{i,j}=0$ or $1$ only (the case $\kappa \to 0$), as opposed to addition over a continuous $u_{t}^{i,j}$ (the case $\kappa \approx 1$). Note also in Fig. \ref{overview2} that, for $\kappa \ge1.9$, inhomogeneities are removed after a transient but the resulting homogeneous state is not completely normalized.

The dynamics of \emph{model II} can be analyzed further using the mean field approximation 
\begin{equation}
u_{t+1}=\mathcal{B}_{\kappa}\left(3-9u_{t},\ \frac{1}{2}\right)+u_{t}\mathcal{B}_{\kappa}\left(4-9u_{t},\ \frac{1}{2}\right) \label{BIlife}
\end{equation}
We describe next the bifurcation diagram of Eq. (\ref{BIlife}) as $\kappa$ is decreased from $\kappa \ge 2$ to $0$: 
\begin{itemize}
\item A bifurcation is encountered at $\kappa \approx 1.95$, which is close to the value $\kappa \approx 1.9$ found in the numerical simulations of the exact dynamics, Eq. (\ref{Lifek}). The system abruptly splits into two branches leading to the bistable regime $A$ (Fig. \ref{BIFli}). Remarkably, the system would normalize when $\kappa \to 0$ only if the lower branch  in Fig. \ref{BIFli} were followed. These facts establish practical limits for restoring and normalization procedures. 
\item A bifurcation of the upper branch is found at $\kappa \approx 1.35$ leading to period-2 oscillations. Further period doubling bifurcations are then observed at $\kappa \approx 1.2$ (as in Fig. \ref{overview}) leading through a period-doubling cascade into chaos which is most prominent at $\kappa=1$ (regime B in Fig. \ref{BIFli}). To substantiate this observation, we have calculated the Lyapunov exponent
\begin{equation}
\lambda (u_{0})\equiv \lim _{{T\to \infty }}{\frac {1}{T}}\sum _{{t=0}}^{{T-1}}\ln \left|\left.\frac{df_{\text{MF}}(u;\kappa)}{du}\right|_{u=u_{t}}\right| 
 \end{equation}
for trajectories of the mean field approximation starting with initial conditions $u_0=0.15$ and $u_{0}=0.45$ in the lower and the upper branches, respectively, and $T=3\cdot 10^4$ (Fig. \ref{lyapu}). While $\lambda(0.15)<0$ for all $\kappa$, we find a positive Lyapunov exponent $\lambda(0.45)>0$ in the range $1.045 \le \kappa \le 1.185$ for a trajectory to the upper branch of the bifurcation diagram. The period doubling bifurcations occur at those $\kappa$ values for which $\lambda(0.45)=0$, consistent with the bifurcation diagram (Fig. \ref{BIFli}) and the numerical simulations of Eq. (\ref{Lifek}), see for example the three last snapshots for $\kappa=1.2$ in Fig. \ref{overview} where period-4 oscillations are observed.

\begin{figure*} 
\begin{center}
\includegraphics[width=0.6 \textwidth]{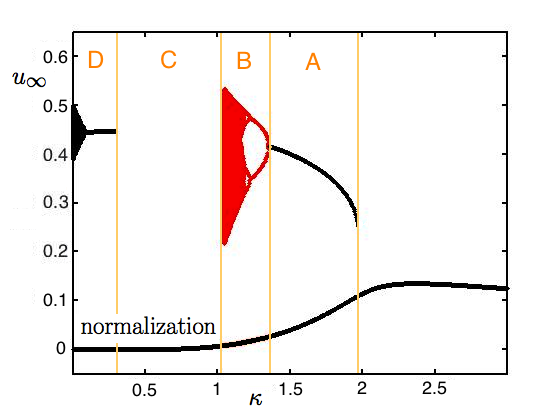}
\caption{\scriptsize{(Color online) Bifurcation diagram calculated from the asymptotic behavior of Eq. (\ref{BIlife}).  The black curves correspond to stationary states $u_{\infty}$ obtained at large times. The light points indicate the period doubling bifurcation cascades into chaos. Note the correspondence of this figure with the results of Figs. \ref{overview} and \ref{overview2}.}} \label{BIFli}
\end{center} 
\end{figure*} 
 
\item In regimes C and D of Fig. \ref{BIFli}, the mean field approximation  fails because it  can no longer be assumed that all neighborhoods are uncoupled and well described by an average cell value. Equation (\ref{Lifek}) needs to be considered in these regimes. Noise is high in regime C (see Figs. \ref{overview} and \ref{overview2} for $\kappa=0.5$) but this noise may have a thermal-like origin (see Ref. \cite{Adachi}). More degrees of freedom may be involved here and it is not  possible to use the mean field approximation,  Eq. (\ref{BIlife}), to account for this dynamics. The results of Fig. \ref{BIFli} clearly show the complex role of the modulating parameter $\kappa$ in the ensemble normalization.
\end{itemize}

The bifurcation diagram (Fig. \ref{BIFli}) also explains the pattern formation in Figs. \ref{overview} and \ref{overview2} for $1\le \kappa \le 1.9$: the upper branch with bifurcations corresponds to the inhomogeneous region and the lower branch to the homogeneous one in Fig. \ref{overview2}. The bifurcation diagram also clarifies why oscillations occur only in the inhomogeneous region.

\begin{figure*} 
\begin{center}
\includegraphics[width=0.6 \textwidth]{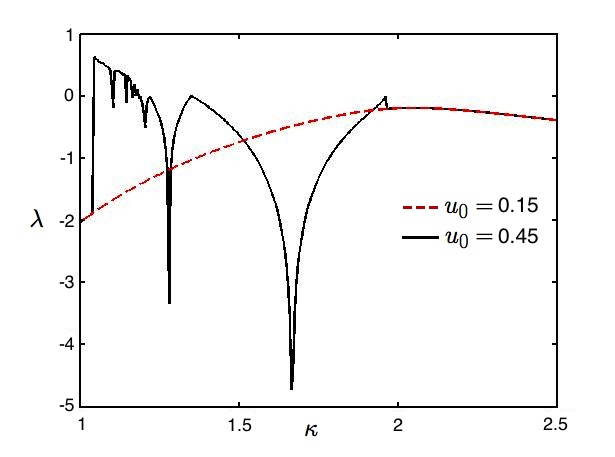}
\caption{\scriptsize{The value of the Lyapunov exponent vs. $\kappa$ for two different initial conditions $u_{0}$.}} \label{lyapu}
\end{center} 
\end{figure*}

An analysis of the noise in the time series of $u_{t}^{i,j}$ for each cell of the ensemble has been carried out in regimes C and D of Fig. \ref{BIFli} (not shown here). The spectrum shifts from uniform noise at $\kappa=1$ to low-frequency ($1/f$) noise at $\kappa \to 0$. The strong correlations found in the limit $\kappa \to 0$, together with the need to take into account local details within a neighborhood, make necessary to use the exact dynamics, Eq. (\ref{Lifek}), instead of its coarse-grained approximation, Eq. (\ref{BIlife}), in that regime. To understand why the mean field approximation breaks down for low $\kappa$ but works well for $\kappa >1$, we note that $\left|\left.\partial f/ \partial s_{t}^{i,j}\right|_{u_{t}^{i,j}=u_{t}}\right| \propto 1/\kappa$ for \emph{model II}, i.e. for $f$ equal to the r.h.s. of Eq. (\ref{Lifek}). Thus, as $\kappa$ is decreased below unity, it is possible to violate the criterion in Eq. (\ref{supercrite}). Conversely, for $\kappa >1$ the contribution of the above derivative is increasingly smaller and Eq. (\ref{supercrite}) can then be satisfied.

Taking together, the different results obtained with \emph{models I} and \emph{II} clearly emphasize the inherent complexity of collective normalization processes based on the restoring of weakened local rules in model multicellular ensembles. Note, in particular, the complex scenario obtained for \emph{model II} when varying the intensity of the intercellular coupling. These results suggest that externally-induced procedures attempting to normalize abnormal cell domains can produce different outcomes depending on the dominant local rules. 

Cells are coupled together and thus their individual properties can be modulated by  ensemble-averaged characteristics such as electric potentials and fields \cite{Cervera4, Adams, Levin1, Levin2}. These characteristics may allow a spatially distributed control of small cellular domains by the conversion of local genetic and bioelectric responses into multicellular states that are regulated by the gap junction interconnectivity. In this context, the approach proposed here should be of interest to different biophysical problems:

1)	Intercellular connectivity is crucial to growth and form. The gap junctions between single cells modulate the rules that instruct pattern regulation \cite{Matthews}. Experimentally, the \emph{functional inhibition} of the gap junctions connecting neighboring cells can be achieved either by injecting a specific factor that targets connexins or by post-translational blocking with an external agent \cite{Emmons}. These processes can be simulated here by \emph{weakening the local rules}. Interestingly, the intercellular gap junctions can contribute to the formation of Turing structures in cortex \cite{Steyn} and are also involved in the bioengineering of excitable tissues capable of information processing \cite{Namara}.

2) Experimentally, addressing gap junctions and connexins as targets in practical problems is difficult because they allow the transmission of a multitude of biochemical and bioelectrical signals between individual cells, which results in a complex context-dependent behavior \cite{Aasen0, Zong}. A limited intercellular communication should enhance autonomous cell behavior and has been related to the initial stages of cancer \cite{Tarin, Sonnenschein, Sonnenschein2, Chernet, Schalper}. However, the outcome to be expected in each experimental case is context-dependent in the sense that it depends not only on the signaling molecule transferred but also on the particular states of the neighboring cells \cite{Aasen0}. Figure 2 suggests that restoring the intercellular coupling might contribute to ensemble normalization but Figs. 5 and 6 show that different responses could also be possible. Taken together, Figs. 3, 5, 6 and 8 provide some qualitative physical insights into this complex problem: the effects of the coupling intensity simulating the intercellular communication here are context-dependent in the sense that the outcomes depend on the local rules and states of the neighboring cells. As it could be expected, Figs. 3 and 8 suggest that a good knowledge of the local rules should facilitate the establishment of appropriate procedures to change the state of cell domains by acting on the intercellular coupling intensity (e.g., by gap junction blockers \cite{Aasen0, Emmons}).

3)	It is possible to analyze the time evolution of multicellular ensembles by direct experimental visualization. For instance, the electrical potential domains formed by cell clusters can be imaged locally by membrane-voltage-reporting dyes \cite{Adams, Levin1, Levin2}. Also, the intercellular coupling may be externally controlled by appropriate agents such as blockers of specific ion channels \cite{Cervera4, Adams, Chernet, Mesnil} and local transfer of microRNAs \cite{Zong}. Weakly coupled map lattices can be of qualitative value to analyze the different spatio-temporal patterns that are obtained in culture assays with multicellular domains.

\section{Conclusions}

The methods used here should have a wide physical significance: they can be applied not only to heterogeneous biological units but also to artificial networks of nanostructures where weak collective coupling may arise because of the individual heterogeneity. Some examples of current interest are nanowire field-effect transistors, nanoparticle-based single electron transistors, and molecular dipoles in monolayers. In these cases, the individual variability results in weak local rules for the system dynamics.

In the case of biological cell networks, theoretical approaches tend to focus on biochemical signals and pathways at the single-cell level. Extensions to tissues are usually based on reaction-diffusion \cite{Meinhardt, Green, Kerszberg} and bioelectrical schemes \cite{Cervera4, Pietak} but network models with different local rules have also been proposed \cite{Bolouri, Torquato, Szabo}. We have shown here that weakly coupled map lattices \cite{JPHYSA} can provide significant insights on intercellular coupling by using two biologically-motivated sets of local rules for the multicellular ensemble dynamics. These rules should be modulated by the protein gap junctions between adjacent cells but the particular mechanisms linking these junctions to processes such as pattern formation and tumorigenesis are not completely known \cite{Matthews}. 

For instance, the bystander effects associated with intercellular coupling may enhance the antitumor effect by transferring specific signaling molecules between neighboring cells \cite{Trosko}. However, the intercellular junctions have context-dependent roles and may show pro- and anti-proliferative effects depending on the particular cell states and the information to be transferred \cite{Aasen0, Emmons}. The rich diversity of results obtained with \emph{models I} and \emph{II} suggests the difficulty of attempting to normalize domains of abnormal cells by restoring weakened local rules: a detailed knowledge of the dominant local rules is necessary to achieve the desired outcomes.

\section{Appendix}
\subsection{The limits $\kappa \to 0$ and $\kappa \to \infty$ of \emph{Model I} yield complementary dynamics}

For $t\ge 1$ and all initial conditions satisfying Eq. (\ref{condition}), the variable $u_{t}^{i,j}$ of \emph{Model I} becomes Boolean in the limits $\kappa \to 0$ and $\kappa \to \infty$. Then, the neighborhood sum $s_{t}^{i,j}$ can only take integer values from 0 to 9, and Eqs. (\ref{limeq0}) and (\ref{limeq1}) reduce, respectively, to
\begin{eqnarray}
u_{t+1}^{i,j}&=&\mathcal{B}\left(9-s_{t}^{i,j},\ \frac{1}{2}\right)=\delta(9-s_{t}^{i,j}) \label{p1} \\
u_{t+1}^{i,j}&=&\mathcal{B}\left(5-s_{t}^{i,j},\ \frac{9}{2}\right)=1-\delta(s_{t}^{i,j}) \label{p2} 
\end{eqnarray}
where
\begin{equation}
\delta(n)={\begin{cases} \text 1 &{\text{if }} n=0 \\ 0 &{\text{if }} n\ne 0   \end{cases}} \label{kronedel}
\end{equation}
is the unit impulse function. Eqs. (\ref{p1}) and (\ref{p2}) are the global complement of each other. That is, the evolutions of $u_{t}^{i,j}$ predicted by these equations are the `negative' of each other under the transformation $\hat{u}_{t}^{i,j}=1-u_{t}^{i,j}$, which exchanges the site states 0 and 1 and transforms the neighborhood sum as $\hat{s}_{t}^{i,j}= 9-s_{t}^{i,j}$. Indeed, inserting Eq. (\ref{p1}) in $\hat{u}_{t+1}^{i,j}= 1-u_{t+1}^{i,j}$ leads to $\hat{u}_{t+1}^{i,j}\equiv 1-\delta(9-s_{t}^{i,j})=1-\delta(\hat{s}_{t}^{i,j})$, the transformed of Eq. (\ref{p2}).

\subsection{Equivalence of Eqs. (\ref{Lifeklim0}) and (\ref{Lifeklim})  when $\kappa \to 0$}

When $u_{t}^{i,j}$ is a Boolean variable, the Game of Life cellular automaton, Eq. (\ref{Lifeklim0}) 
\begin{equation}
u_{t+1}^{i,j}={\begin{cases} u_{t}^{i,j} &{\text{if }} s_{t}^{i,j}-u_{t}^{i,j}=2 \\ 1 &{\text{if }} s_{t}^{i,j}-u_{t}^{i,j}=3    \\ 0&{\text{otherwise }} \end{cases}} \label{Lifeklim0p}
\end{equation}
can be written in terms of the unit impulse function as
\begin{eqnarray}
u_{t+1}^{i,j}&=&\delta(s_{t}^{i,j}-u_{t}^{i,j}-3)+u_{t}^{i,j}\delta(s_{t}^{i,j}-u_{t}^{i,j}-2)=\delta(s_{t}^{i,j}-3)+u_{t}^{i,j}\delta(s_{t}^{i,j}-4)
\label{Lifle}
\end{eqnarray}
which is equivalent to Eq. (\ref{Lifeklim})
\begin{equation}
u_{t+1}^{i,j}=\mathcal{B}\left(s_{t}^{i,j}-3, \frac{1}{2}\right)+u_{t}^{i,j}\mathcal{B}\left(s_{t}^{i,j}-4, \frac{1}{2}\right)={\begin{cases} \text 1 &{\text{if }} s_{t}^{i,j}=3 \\ u_{t}^{i,j} &{\text{if }} s_{t}^{i,j}=4    \\ 0&{\text{otherwise }} \end{cases}} \label{Lifeklimp}
\end{equation}

\section*{Acknowledgements}
Financial support by the Spanish Ministry of Economic Affairs and Competitiveness (MAT2015-65011-P), and FEDER are acknowledged. We are grateful to the anonymous referees for helpful suggestions. This paper is dedicated to the memory of Prof. Juan de la Rubia.

\end{document}